\documentstyle[12pt,epsf,epsfig]{article}
\def\be{\begin{equation}}
\def\ee{\end{equation}}
\def\ba{\begin{eqnarray}}
\def\ea{\end{eqnarray}}

\def\bdm{\begin{displaymath}}
\def\edm{\end{displaymath}}
\def\la{~\mbox{\raisebox{-.6ex}{$\stackrel{<}{\sim}$}}~}
\def\ga{~\mbox{\raisebox{-.6ex}{$\stackrel{>}{\sim}$}}~}
\def\bq{\begin{quote}}
\def\eq{\end{quote}}

 at 10truept

\newcommand{\beq}{\begin{equation}}
\newcommand{\eeq}{\end{equation}}
\newcommand{\beqa}{\begin{eqnarray}}
\newcommand{\eeqa}{\end{eqnarray}}

\def\la{~\mbox{\raisebox{-.6ex}{$\stackrel{<}{\sim}$}}~}
\def\ga{~\mbox{\raisebox{-.6ex}{$\stackrel{>}{\sim}$}}~}

\def\ltap{\ \raise.3ex\hbox{$<$\kern-.75em\lower1ex\hbox{$\sim$}}\ }
\def\gtap{\ \raise.3ex\hbox{$>$\kern-.75em\lower1ex\hbox{$\sim$}}\ }
\def\gl{\ \raise.5ex\hbox{$>$}\kern-.8em\lower.5ex\hbox{$<$}\ }
\def\roughly#1{\raise.3ex\hbox{$#1$\kern-.75em\lower1ex\hbox{$\sim$}}}

\begin{document}

\thispagestyle{empty}
\begin{flushright}
arXiv:0711.3210 [hep-th]\\ November 2007
\end{flushright}
\vspace*{.8cm}
\begin{center}
{\Large \bf Brane Induced Gravity: Codimension-2\footnote{Based on the talks given at the
``Sowers Workshop", Virginia Tech, May 14-18, 2007, ``Cosmology
and Strings" workshop at ICTP, Trieste, Italy, July 9-13, 2007,
``Dark Energy In the Universe", Hakone, Japan, Sep 1-4, 2007 and
``Zagreb Workshop 2007", Zagreb, Croatia, Nov 9-11, 2007.}
}\\

\vspace*{1.5cm} {\large Nemanja Kaloper\footnote{\tt
kaloper@physics.ucdavis.edu}}\\
\vspace{.5cm} {\em Department of Physics, University of
California, Davis,
CA 95616}\\
\vspace{.15cm} \vspace{1.2cm} ABSTRACT
\end{center}

We review the results of arXiv:hep-th/0703190, on brane
induced gravity (BIG) in $6D$. Among a large diversity of
regulated codimension-2 branes, we find that for near-critical
tensions branes live inside very deep throats which efficiently
compactify the angular dimension. In there, $4D$ gravity first
changes to $5D$, and only later to $6D$. The crossover from $4D$
to $5D$ is independent of the tension, but the crossover from $5D$
to $6D$ is not. This shows how the vacuum energy problem manifests
in BIG: instead of tuning vacuum energy to
adjust the $4D$ curvature, generically one must tune it to get the
desired crossover scales and the hierarchy between the scales
governing the $4D \rightarrow 5D \rightarrow 6D$ transitions. In
the near-critical limit, linearized perturbation theory remains
under control below the crossover scale, and we find that
linearized gravity around the vacuum looks like a scalar-tensor
theory.

\vfill \setcounter{page}{0} \setcounter{footnote}{0}
\newpage

\section{Introduction}

Cosmological constant problem is the deepest problem of
fundamental physics. To date, the attempts to explain a small
cosmological constant in effective field theory coupled to General
Relativity using dynamical adjustment have not yielded an answer
\cite{Wein}. The perspective may change if our universe is a brane
in extra dimensions \cite{rusha}-\cite{selftun}. In such setups
one may divert the brane vacuum energy into the extrinsic
curvature \cite{selftun}, so that the vacuum energy could remain
invisible to long distance $4D$ gravity, yielding a low curvature
$4D$ geometry. However, while recovering exact $4D$ General
Relativity at large distances, by using additional branes in the
bulk one rediscovers a fine-tuning similar to the standard $4D$
one \cite{selftun,nillest}. A program of \cite{msledcc} has been initiated to
seek the means to relax these fine-tunings in codimension-2 brane setups.
Interesting precursors to these ideas however can be found already in 
\cite{markus}.

One may try to alleviate the problem by only approximating $4D$
General Relativity at large distances, as in the brane induced
gravity (BIG {\it a.k.a.} DGP) \cite{DGP}. There the extra-dimensional space may
have an infinite volume, so that the $4D$ graviton completely
decouples. Thus $4D$ gravity must emerge from the exchange of the
continuum of bulk modes. This may occur with the help of the
induced curvature terms on the brane \cite{DGP}, which control the
momentum transfer due to the scattering of virtual bulk gravitons
and may yield a force which scales as $1/r^{2}$, simulating $4D$
behavior \cite{DGP}. The exact gravitational shock waves of
\cite{shocks} confirm this nicely.

These mechanisms may operate on higher-codimension defects
\cite{DGP,highercod}. In \cite{kkland}, with D. Kiley we explored
in a very detailed fashion the BIG mechanism on general
codimension-2 branes floating in $6D$.  We presented exact
background solutions with flat $4D$ geometries for nonvanishing
tensions, and regulating the short distance properties by
considering hollow cylindrical branes we calculated the various
regimes of gravity. The most interesting case, where the theory
displayed a remarkably good perturbative behavior, was for
near-critical tensions, where the deficit angle around the brane
was almost $2\pi$. In this limit the bulk compactifies  to a very
thin sliver \cite{kalwall}, inside which bulk gravity is $5D$.
Because of this gravity along the brane changes dimensions with
distance according to $4D \rightarrow 5D \rightarrow 6D$. The
crossover scale where gravity stops being $4D$ is independent of
the tension, and can be quickly obtained by using the naive
crossover formula on a codimension-1 brane, realized as a wrapped
4-brane in a $6D$ flat space with a compact circle: by Gauss
formulas for Planck masses $M_{4 \, eff}^2 = M_5^3 r_0$ and $M_{5
\, eff}^3 = M_6^4 r_0$, we find exactly $r_c \sim \frac{M_{4 \,
eff}^2}{M_{5 \, eff}^3} \sim \frac{M^3_5}{M^4_{6 }}$, which is our
crossover formula from the exact shock waves, and which is the
same as the see-saw scale $r_c \sim \frac{M^2_4}{M^4_{6}r_0}$ of
\cite{highercod}. Thus we see precisely {\it how} the see-saw
mechanism emerges. However, BIG is not free of the vacuum energy
problem: although the brane is flat, when we fix $M_4$, we must
finely tune the tension of the brane to get the bulk sliver
geometry, that yields the separation of the $4D$ and $6D$ regimes
by a $5D$ one.

From the shock waves we have seen that the theory contains $4D$
General Relativity. We unveil the rest in the linearized
perturbation theory, which shows that there are scalar modes with
gravitational couplings to matter. In the near-critical limit
linearized perturbation theory around the vacuum remains under
control in the $4D$ regime below the crossover scale. While the
radion is suppressed, the helicity-0 mode mediates an extra long
range force around the $4D$ vacuum backgrounds, and where the
linearized theory is $4D$ it a Brans-Dicke gravity with
$\omega=0$. This would disagree with the classic tests of General
Relativity \cite{vdvz}, but perhaps non-linearities or dynamics on
vacua of more complex structure could come to the rescue. As the
tension decreases towards sub-critical values the scalars start to
couple strongly, and the perturbativity of the $4D$ regime is lost
exactly at the $4D$ crossover scale $r_c$, which is the Vainshtein
scale \cite{veinsh} of the vacuum itself. This happens since the
scalar gravitons probe the regulating axion flux, which sources
brane bending. On the other hand, we do not find any instabilities
in the leading order of perturbation theory.

In this note, we will review the near-critical limits for the sake
of clarity. The detailed analysis of the general sub-critical
setups with flat $4D$ geometries can be found in \cite{kkland}. We
should note that very recently in \cite{many} similar
features of BIG theory in $6D$ environments were argued to
emerge in a framework  with more branes. Also, some 
ideas along these lines were explored in a setup 
with codimension-3 branes with a solid deficit angle in \cite{gia}.

\section{Regulated Near-critical Branes}

We model a 3-brane in $6D$ (codimension-2) by a $4$-brane wrapped
as a cylinder \cite{msled,pesota}, in order to regulate the short
distance singularities which will appear even in the classical
theory. When the 4-brane has nonzero tension, with the vacuum
action $S_{\rm vacuum} = - \int d^5 x \sqrt{g_5} \lambda_5$, its
stress energy tensor will be $T^A{}_B = - \lambda_5 \delta^{}(\rho
- r_0) \, {\rm diag}(1,1,1,1,1,0)$, where we wound it around the
circle of radius $r_0$, with $\rho$ along the normal to the brane.
Thus, to wrap the 4-brane into a cylinder, we must cancel the
pressure $\propto \lambda_5$ in the compact direction. A simple
way to do it is to put an axion-like field $\Sigma$ on the
4-brane, with a vacuum action $S_{vacuum} = - \int d^5 x
\sqrt{g_5}\big(\lambda_5 + \frac{1}{2} g^{ab}\partial_a \Sigma
\partial_b \Sigma\big)$. Taking the fourth coordinate to be the
angle on the compact circle $\phi$ and substituting the
Scherk-Schwarz ansatz $\Sigma = q \phi$ we choose $q$ to precisely
cancel $T^\phi{}_\phi$. This requires
\be \lambda_5 = \frac{1}{2}q^2 g^{\phi \phi} \, , \label{tuning}
\ee
where $g^{\phi\phi} = \frac{1}{r_0^2}$ is the inverse radius
squared of the compact dimensions. Then the remaining components
of the 4-brane stress energy become precisely $T^\mu{}_\nu = -2
\lambda_5 \delta^\mu{}_\nu$. Thus the brane source now reads
$T^A{}_B = - 2\lambda_5 \delta(\rho - r_0) \, {\rm diag}(1, 1, 1,
1, 0, 0)$, and the tensor structure is precisely the same as in
the stress energy tensor of a thin 3-brane. By inspection, we see
that if we shift the argument of the remaining radial
$\delta$-function to $\rho - r_0$, it becomes $\delta^{(2)}_{\rm
thick}(\vec y) = \frac{1}{2\pi r_0} \delta(\rho - r_0)$, and so
the effective $4D$ tension is
\be
\lambda = 4\pi r_0 \lambda_5 \, ,
\label{lambdas}
\ee
relating the $5D$ and the effective $4D$ vacuum energies. The
effective $4D$ vacuum energy $\lambda$ contains the contributions
from the classical axion field $\Sigma$, doubling up its value,
because of the cancellation condition of Eq. (\ref{tuning}).

\begin{figure}[thb]
\vskip.6cm
\centerline{\includegraphics[width=0.35\hsize,width=0.35\vsize,angle=0]{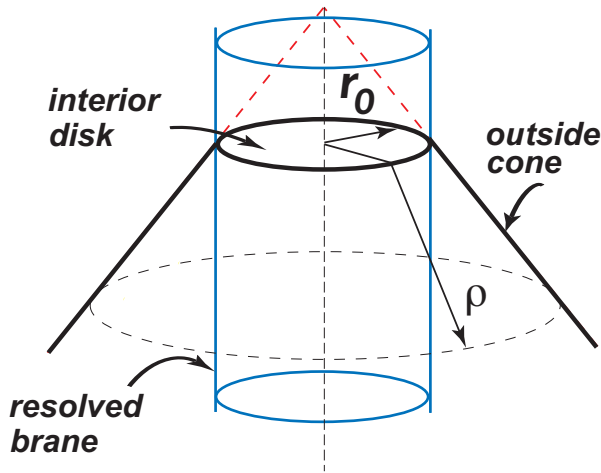}}
\caption{{$2D$ bulk geometry of the resolved brane.}
\label{figure}} \vskip.5cm
\end{figure}

The field equations for a 4-brane in a $6D$ bulk which include
brane localized gravity terms are \cite{DGP,highercod}, in
Gaussian-normal gauge, and with the 4-brane residing at $\rho =
r_0$,
\be M_6^4 G_6{}^A{}_B + M_5^3 G_5{}^a{}_b \delta^A{}_a
\delta^b{}_B \delta(\rho-r_0) =  T^a{}_b \delta^A{}_a
\delta^b{}_B  \delta(\rho-r_0) \, ,
\label{thickfieldeqs}
\ee
which we apply to our cylindrical brane vacuum, imposing that the
stress energy tensor is covariantly conserved, that requires
$\partial^a \partial_a \Sigma = 0$. It is easy to check that by
axial symmetry the ansatz $\Sigma = q \phi$ trivially solves the
latter equation. Then, the tensor structure of the source,
$T^A{}_B = - 2\lambda_5 \delta(\rho-r_0) \, {\rm diag}(1, 1, 1, 1,
0, 0)$, guarantees that the flat $4D$ metric is a solution, if the
tension is again off-loaded into the bulk, like in the thin
3-brane case \cite{raman}. Tracing the field equations
(\ref{thickfieldeqs}), and using (\ref{tuning}) we find that the
condition for this is that the metric in the remaining two
dimensions, coordinatized by the angular direction along the brane
and the radial distance away from it has curvature
\be
R_2 = \frac{4\lambda_5}{M_6^4}\delta(\rho-r_0) \, .
\label{twodthick}
\ee
It is straightforward to find the metric \cite{kkland}, which is
\be ds_6{}^2 = \eta_{\mu\nu}dx^\mu dx^\nu + d\rho^2 +\bigg[
(1-b\Theta(\rho-r_0))\rho + b r_0 \Theta(\rho-r_0)\bigg]^2 d\phi^2 \, ,
\label{thick6dvac} \ee
where $b = \frac{2 \lambda_5 r_0}{M_6^4}$. For $\rho < r_0$ the
$2D$ part of the metric is $ds_2{}^2 = d\rho^2 +\rho^2 d\phi^2$,
i.e. a flat disk, while for $\rho > r_0$ it becomes $ds_2{}^2 =
d\rho^2 + \big((1-b)\rho+b\,r_0\big)^2 d\phi^2 = d\rho^2 +
(1-b)^2\big(\rho+\frac{b}{1-b}\,r_0\big)^2 d\phi^2$, precisely the
metric on a cone, which we see by comparing it to the thin 3-brane
metric from the previous section. Thus the combined solution
represents a truncated cone, with a flat mesa of radius $r_0$ on
top, as depicted in Fig. (\ref{figure}). The geometry of the brane
is $4D$ Minkowski $\times$ circle.

\begin{figure}[thb]
\vskip.6cm
\centerline{\includegraphics[width=0.3\hsize,width=0.3\vsize,angle=0]{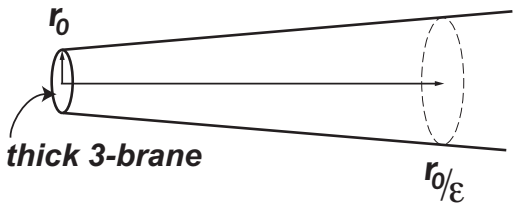}}
\caption{{$2D$ bulk geometry of the near-critical resolved
3-brane.} \label{sliver}} \vskip.5cm
\end{figure}

Looking at Eq. (\ref{thick6dvac}) for $\rho > r_0$ we see that the
critical string tension, where the deficit angle becomes $2\pi$,
is again given by $b=1$, which now corresponds to the tension
$\lambda_{5 \, cr} = \frac{M_6^4}{2r_0}$. Comparing to the
relation between the $5D$ tension $\lambda_5$ and the effective
$4D$ tension $\lambda$, given by Eq. (\ref{lambdas}) we find that
the critical value of the effective $4D$ tension is, not
surprisingly, $\lambda_{cr} = 2\pi M_6^4$. Now, when $b =
1-\epsilon$, it's easy to see that the bulk cone looks like a
sliver. In this limit, the exterior bulk metric is approximately
$ds_6{}^2 = \eta_{\mu\nu}dx^\mu dx^\nu + d\rho^2 + \big(\epsilon
(\rho - r_0)\,+\,r_0\big)^2\ d\phi^2$. This looks like the metric
on a cylinder for $r_0 \le \rho \la r_0/\epsilon$, because the
variation of the radius of the sliver, due to the change of radial
distance $\rho$, is very small compared to the radius of the brane
$r_0$. Near-critical branes reside inside deep bulk throats that
only asymptotically open into conical geometries \cite{kalwall},
see Fig (\ref{sliver}).

Note, that our construction is completely covariant from the bulk
point of view, since we can define the regulator by a variational
principle, from a well defined 4-brane action. Therefore we can
explicitly check the properties of brane localized gravity, and
calculate the crossover scale inferred in \cite{highercod}
exactly.

\section{Matter on Thick Branes}

We can take the matter to be localized on the 4-brane, and
described by the action $S_{\rm matter} = - \int d^5 x \sqrt{g_5}
\, {\cal L}_{\rm matter}$. Matter couples to the induced metric on
the 4-brane $g_{5 \, ab}$. Since we are interested in the behavior
of the theory at length scales greater than the radius of the
brane, $r \gg r_0$, we can dimensionally reduce the
brane-localized sector on this direction, representing all brane
fields by the standard Fourier expansion $\Phi_{\cal N} (x^\mu,
r_0 \phi) = \sum^\infty_{n = - \infty} \Phi_{{\cal N}, n}(x^\mu)
\, e^{i n \phi}$, where ${\cal N}$ represents the quantum numbers
of any particular representation on the brane. The Fourier
coefficients then give rise to the $4D$ KK towers of fields, with
masses $M^2 = m^2 + n^2/r_0^2$, where $m$ are the explicit $5D$
masses and the mass gap is $1/r_0$. This expansion also applies to
the brane induced metric and its curvature. Thus, at distances $r
\gg r_0$ we can always model the angular dependence of any
configuration  as a small, Yukawa-suppressed correction from a KK
momentum mode stretching along the compact circle.

To determine the stress energy tensors of the thin loops of matter
on the wrapped brane we can model them by a Nambu-Goto action for
a string, $S_{\rm matter} = -\mu \int d^2\sigma \sqrt{\gamma} $,
where $\mu$ is the mass per unit length of the string,
$\sigma^\alpha$ are its two worldsheet coordinates, which we will
gauge fix to be $t$ and $r_0 \phi$, and $\gamma_{\alpha\beta}$ is
the induced worldsheet metric. We then vary this action with
respect to the metric of the target space on which the string
moves, in this case the metric of the background thick brane.
Using the standard definition of the stress energy tensor
$\tau^{ab}$, given by the functional derivative equation $\delta
S_{\rm matter} = \frac{1}{2}\int d^5x\,\sqrt{g_5} \delta g_{ab}
\tau^{ab}$, to read off the stress energy tensor of the loop.
Rewriting then the action as $S_m = -\mu \int d^2\sigma d^5 x
\,\sqrt{\gamma}\, \delta^{(5)}(x^M - x^M(\sigma))$, where the
$\delta$-function puts the string on its shell, with the gauge
choice $\sigma^0 = x^0$ and $\sigma^1 = r_0 \phi$, and using the
vacuum solution (\ref{thick6dvac}) as the target space, so that
$\gamma_{\alpha\beta} = \eta_{\alpha\beta}$, we find
\be \tau^{ab} = -\mu \int d^2\sigma \delta^{(5)}(x^c -
x^c(\sigma)) \eta^{\alpha \beta}\frac{\partial x^a}{\partial
\sigma^\alpha}\frac{\partial x^b}{\partial \sigma^\beta}.\\
\label{eqn:stringseint}
\ee
This string stress energy is conserved, $\nabla_a \tau^{ab} = 0$,
by virtue of the string equations of motion. For static sources,
the string representing only the lightest modes must be
translationally invariant in the compact direction, because by the
discussion above any inhomogeneities along the string can be
viewed as heavy KK states at distances $r \gg r_0$, which
decouple. For such axially symmetric sources, the stress energy
boils down to
\be \tau^a{}_b = -\mu \, \delta^{(3)}(\vec{x}) \,
\textrm{diag}(1,0,0,0,1) \, . \label{eqn:strinsediag} \ee
The parameter $\mu$ is the total  rest mass per unit length of the
configuration. For a relativistic string, we can simply boost
(\ref{eqn:strinsediag}) along the direction of motion
$x_\parallel$, go to the light cone coordinates  $x_\parallel =
v+u$, $t=v-u$ and take the limit of infinite boost parameter while
sending $\mu \rightarrow 0$ \cite{aichsexl,thooft}. Since the
momentum of the string $p$ is the length of the string $2\pi r_0$
multiplied by the finite limit of $\mu \cosh \beta$ as $\beta$
diverges, for sources composed of the lightest modes as in
(\ref{eqn:strinsediag}) this yields, thanks to the properties of
the vacuum (\ref{thick6dvac}),
\be \tau^\mu{}_\nu = \frac{2p}{2\pi r_0 \sqrt{g_4}}g_{4uv}
\delta(u)\delta^{(2)}(\vec{x}_\bot)\delta^\mu{}_v \delta^u{}_\nu
\, . ~~~~~~~~~~~~~ \tau^\phi{}_\phi = 0 \, , \label{shocktmunu}
\ee

To find the gravitational fields of these sources, we need to
solve the BIG field equations
(\ref{thickfieldeqs}), in Gaussian-normal gauge with the brane at
$y=\rho-r_0 = 0$ and with a specified $T^a{}_b$, which includes
the background tension and axion flux, and $\tau^a{}_b$ as given
in (\ref{eqn:strinsediag}) or (\ref{shocktmunu}). Rewriting field
equations as the conditions on the Ricci tensor, we need to solve
\be M_6^4 R_6{}^A{}_B = \bigg[\bigg(T^a{}_b - M_5^3
\big(R_5{}^a{}_b - \frac{1}{2}\delta^a{}_b
R_5\big)\bigg)\delta^A{}_a \delta^b{}_B -\frac{1}{4}\delta^A{}_B
\big(T^a{}_a + \frac{3}{2}M_5^3 R_5\big)\bigg] \delta(\rho-r_0) \,
. \label{riccieq} \ee

\section{Shocks}

We determine the shock wave solution by introducing a
discontinuity in the metric of Eq. (\ref{thick6dvac}) according to
the prescription of \cite{thooft,shocks}, and demanding that the
wave profile solves Eq. (\ref{riccieq}). By axial symmetry of the
source, we can only look for shock wave profiles
$f(\vec{x}_\bot,\rho)\delta(u)$ that depend on the transverse
coordinates along the brane $\vec{x}_\bot$ and the transverse
distance $\rho$ orthogonal to the brane. Thus the shock wave
metric is
\be ds_6{}^2 = 4dudv - 4 \delta(u) f du^2 + d\vec{x}^2_\bot +
d\rho^2 + \bigg[(1-b\,\Theta(\rho-r_0))\rho+b\,r_0 \Theta(\rho
-r_0)\bigg]^2 d\phi^2 \, . \label{shockthick} \ee
Substituting this and (\ref{shocktmunu}) in Eq. (\ref{riccieq}),
we find shock wave equation
\be \nabla_4{}^2 f + \frac{M_5^3}{M_6^4} \nabla_2{}^2 f \,
\delta(\rho- r_0 ) = \frac{2p}{2\pi r_0 M_6^4} \,
\delta^2(\vec{x}_\bot) \, \delta(\rho-r_0) \, . \label{lapeom} \ee
The Laplacian $\nabla_4{}^2$
is defined on the truncated cone (\ref{thick6dvac}), such that for
axially symmetric shocks, $\nabla_4{}^2 f = \nabla_{\vec
x_\bot}{}^2 f
+ f'' +\frac{1-b\Theta(\rho-r_0)}{(1-b\Theta(\rho-r_0))\rho
+ br_0\Theta(\rho-r_0)} f'$, where
the prime denotes the $\rho$-derivative. The latter term contains
a jump across the 4-brane.

To solve (\ref{lapeom}), we Fourier transform in the brane
transverse space. This yields \cite{kkland}
\be \big(\nabla_{\vec{y}}^2 - k^2\big)\varphi_k =
\bigg(\frac{2p}{2\pi r_0 M_6^4} + \frac{M_5^3}{M_6^4} k^2
\varphi_k \bigg) \delta(\rho- r_0) \, . \label{varphithick} \ee
We can use the standard methods for finding bulk field wave
functions in the codimension-1 brane setups: we solve Eq.
(\ref{varphithick}) inside $(-)$ and outside $(+)$ the brane, and
then use the boundary conditions, which are the continuity of
$\varphi_k$, the jump of $\varphi_k'$ as prescribed by Gaussian
pillbox integration of (\ref{varphithick}) around the brane, and
the regularity of the solution in the center of (\ref{thick6dvac})
and at infinity. Then using the axial symmetry in the $\vec
x_\bot$ plane to integrate over the transverse spatial angle in
the Fourier transform formula, which gives $f =
\frac{1}{2\pi}\int_0^\infty dk \, k \varphi_k
J_0(k|\vec{x}_\bot|)$, we finally obtain the expression for the
shock wave profile \cite{kkland}:
\be f(\vec x_\bot, \rho)  = -\frac{p}{2\pi^2 r_0} \int_0^\infty dk
\, \frac{I_0(k\rho_<) K_0(k\rho_>) J_0(k|\vec{x}_\bot|)}{M_6^4
\Bigl(I_0(kr_0) K_1(k\frac{r_0}{1-b}) + I_1(kr_0)
K_0(k\frac{r_0}{1-b})\Bigr) +M_5^3 k
I_0(kr_0)K_0(k\frac{r_0}{1-b}) } , \label{thickshosol1} \ee
where
\be I_0(k\rho_<) K_0(k\rho_>)  = \cases { I_0\Bigl(k\rho\Bigr)
K_0\Bigl(k\frac{r_0}{1-b}\Bigr) \, ,   & ~~~~ $\rho \le r_0$ \, ;
\cr I_0\Bigl(k r_0 \Bigr)  K_0\Bigl(k(\rho +
\frac{br_0}{1-b})\Bigr) \, ,  & ~~~~ $\rho \ge r_0$ \, . }
\label{thickshosol2} \ee
This solution is clearly regular on and off the brane,
as long as $r_0 > 0$. On the brane,
$\rho = r_0$, it reduces to
\be f(\vec x_\bot, r_0)  = -\frac{p}{2\pi^2 r_0} \int_0^\infty dk
\, \frac{I_0(kr_0) K_0(k\frac{r_0}{1-b})
J_0(k|\vec{x}_\bot|)}{M_6^4 \Bigl(I_0(kr_0) K_1(k\frac{r_0}{1-b})
+ I_1(kr_0) K_0(k\frac{r_0}{1-b})\Bigr) +M_5^3 k
I_0(kr_0)K_0(k\frac{r_0}{1-b}) } \, , \label{thickshosol1br} \ee
where all the individual contributions remain finite.

\section{Scales of Gravity}

Now we can look at the field along the brane,
(\ref{thickshosol1br}). The solution is dominated by the modes
with the momenta $k \sim 1/|\vec x_\bot |$, due to the oscillatory
nature of $J_0$. The contributions of the modes with momenta  $k$
far from $1/|\vec x_\bot |$ will interfere destructively. At
transverse distances $|\vec x_\bot| \gg r_0$, we need to focus on
the momenta for which $k r_0 \ll 1$. Thus we can always replace
the terms $\propto I_n(kr_0)$ by their small argument expansion.
For sub-critical branes, $1-b \sim {\cal O}(1)$, and we can
likewise replace any $K_n(k\frac{r_0}{1-b})$ by their small
argument expansion too. However, in the near-critical limit the
deficit angle approaches $2\pi$ and so $|1-b| \ll 1$. Hence the
argument of $K_0(k\frac{r_0}{1-b})$ may be very large even when
$kr_0 \ll 1$. Thus we must consider the near-critical branes very
carefully.

For sub-critical branes, $1-b \la {\cal O}(1)$, $kr_0 \ll 1-b$,
and we can approximate all Bessel functions by their small
argument form. The shock along the brane then becomes
\be f(\vec x_\bot, r_0) \simeq -\frac{p}{2\pi^2
M_6^4}\int_0^\infty dk\, k \,
\frac{\ln\Bigl[\frac{2(1-b)}{kr_0}\Bigr]
J_0(k|\vec{x}_\bot|)}{(1-b)+\frac{M_5^3 r_0}{M_6^4} \, k^2 \,
\ln\Bigl[\frac{2(1-b)}{kr_0}\Bigr]} \, . \label{approxshock} \ee
When $k \sim 1/|\vec x_\bot|$ is large enough so that the second term in
the denominator dominates, which
happens when
$k^2 > k^{2}_c(k) =\frac{ (1-b) M_6^4 }{M_5^3 r_0
\ln\Bigl[\frac{2(1-b)}{kr_0}\Bigr]}$, the shock wave profile is
\be  f(\vec x_\bot, r_0) \simeq -\frac{p}{2\pi^2 M_5^3 r_0}
\int_{k_c}^\infty dk \frac{J_0(k|\vec{x}_\bot|)}{k} = \frac{p}{\pi
M_4^2} \ln\bigg(k_c |\vec{x}_\bot|\bigg) + \ldots \, .
\label{r0limit} \ee
where we have used Gauss law to normalize the lightest KK modes in
the expansion of the brane induced curvature on the circle of
radius $r_0$ to $M_4^2 = 2\pi M_5^3 r_0$. This is  precisely the
$4D$ Aichelburg-Sexl solution \cite{aichsexl}, with the log
profile normalized to the inverse of the critical momentum $k_c$.
So for sub-critical branes the gravitational field looks $4D$,
like (\ref{r0limit}), only when $r < r_c = 1/k_c$. At distances
larger than $r_c$, gravity eventually changes to $6D$. In this
limit (\ref{thickshosol2}) or (\ref{thickshosol1br}) yield
\be f \simeq -\frac{p}{2\pi^2(1-b)M_6^4}\int_0^{\infty} dk \,k\,
I_0(kr_0)K_0(k(\rho+\frac{r_0}{1-b})) J_0(k|\vec{x}_\bot|) +
\ldots \, , \label{eqn:f1} \ee
where the integral can be done in closed form, leading to
\cite{kalkil}
\be f \simeq -\frac{p}{2\pi^2(1-b)M_6^4} \frac{1}{|\vec{x}_\bot|^2
+ \rho^2} + \ldots \, . \label{venez} \ee
This is the $6D$ gravitational shock wave, constructed in
\cite{higherdwaves}, with the `lightning rod' term accounting for
the brane tension as found in \cite{kalkil}. The distance scale
which controls the transition, $r_c^2(k) = \frac{M_4^2}{2\pi (1-b)
M_6^4} \ln\Bigl[\frac{2(1-b)}{kr_0}\Bigr]$, is {\it very
different} from the conjectured see-saw scale of \cite{highercod}.
So for sub-critical branes the $4D \rightarrow 6D$ transition is
controlled by the `naive' ratio of brane and effective bulk Planck
scales, and depends on the UV cutoff $r_0$ of the brane core only
through the logarithmic correction reminiscent of `running' in
real space, which was discussed for locally localized gravity in
\cite{kalsorbo}.

In the near-critical limit, $b \rightarrow 1$, the story is a lot
more interesting. In this case the bulk around the brane
compactifies to a long cylinder, which opens up into a cone only
very far from the brane \cite{kalwall}. Bulk gravity in the throat
is in the $5D$ regime, which separates the $4D$ and $6D$ ones.
Thus the crossover scale beyond which gravity is not $4D$
determines where it changes to $5D$ gravity which lives inside the
throat. In the shock wave solution (\ref{thickshosol1br}), we can
always use small argument expansion for $I_n(z)$ at large
distances from the source, but in the functions $K_n(z)$ we must
use the large argument expansion for the momenta $1-b \ll kr_0 \ll
1$, which correspond to length scales $\ell \la \frac{r_0}{1-b}$
inside the conical throat. In this case gravity is $4D$ for
momenta $k > k_c =
\frac{M_6^4}{M_5^3}\frac{K_1(k\frac{r_0}{1-b})}{K_0(k\frac{r_0}{1-b})}$,
which thanks to the limit $K_1 \rightarrow K_0$ saturates to $k_c
= \frac{M_6^4}{M_5^3}$. Therefore the scale below which gravity is
$4D$ is $r_c = \frac{M_5^3}{M_6^4}$, which is valid as long as
$r_c < \frac{r_0}{1-b}$. This scale is exactly the see-saw scale
of \cite{highercod}: using Gauss's law $M_4^2 = 2\pi M^3_5 r_0$,
we find
\be r_c = \frac{M_4^2}{2\pi M_6^4 r_0} \, . \label{seesaw} \ee
This explains the see-saw effect, showing that at least with our
regularization, it does happen naturally but only in the
near-critical limit.

To confirm that inside the conical throat gravity is $5D$, we can
look at our solution for a fixed distance $|\vec x_\bot|$ and take
the limit $b \rightarrow 1$ while holding $r_0$ fixed. Replacing
$K_n$'s by their large argument expansion for any finite momentum,
cancelling the like terms, and taking the thin brane limit, we
find \cite{kkland}
\be f(\vec x_\bot, \rho)  = -\frac{p}{2\pi^2} \int_0^\infty dk \,
\frac{ J_0(k|\vec{x}_\bot|)}{M_6^4r_0  +  M_5^3 r_0 k } = - 2 p
\int_{-\infty}^\infty \frac{d^2\vec k}{(2\pi)^2} \, \frac{e^{i
\vec k \cdot \vec x_\bot}}{2\pi M_6^4r_0 \, k +  2\pi M_5^3 r_0 \,
k^2 } \, . \label{5dthickshosol1} \ee
This is {\it precisely} the shock wave profile on a flat thin
3-brane in $5D$, on the normal branch, with brane and bulk Planck
masses $M_{4 \, eff}^2 = 2\pi M^3_5 r_0$ and $M_{5 \, eff}^3 =
2\pi M^4_6 r_0$ \cite{shocks,kkland}.

So the crossover scale for sub-critical and near-critical branes,
beyond which gravity is not $4D$ anymore, is given by
\be
r^2_c = \cases{ \frac{M_4^2}{2\pi (1-b) M_6^4}
\ln\Bigl[\frac{2(1-b)}{kr_0}\Bigr] \, ,
& \, \,  ${\rm subcritical ~tension}$; \cr
\frac{M_4^4}{4\pi^2 M_6^8 r_0^2} \, ,
& \, \,  ${\rm near-critical ~tension}$, }
\label{crosssum}
\ee
only in the latter case agreeing with the see-saw scale discussed
in \cite{highercod}.

\section{Gravity of Slow Sources}

The shock wave analysis can only discover the presence of
helicity-2 modes, since in the limit of relativistic sources when
the rest masses vanish, the source stress energy tensor becomes
traceless. Such sources will not excite the additional helicities
in the graviton spectrum in the linearized theory. Thus to see
what else is in the spectrum, one eventually needs to engage in
the laborious procedure of gravitational perturbation theory
around slow, nonrelativistic, sources. This analysis has been done
in full detail in \cite{kkland}. Here we will merely recall few
basic results.

The perturbed metric in the brane-fixed Gaussian-normal gauge,
\be ds_6{}^2 = (\eta_{\mu\nu} +  h_{\mu\nu}) \, dx^\mu dx^\nu +
d\rho^2 + (1+{\tilde \Phi})\,\alpha(\rho) \, d\phi^2  \, ,
\label{gngbfix} \ee
with $\tilde \xi = 0$ and $\alpha = g_{\phi\phi}$ of the
background. The axion perturbation obeys $\partial_4{}^2 \sigma =
0$, and so in the linear order the axion is not sourced by axially
symmetric matter distributions and can be set to zero.
Substituting this in the field equations (\ref{riccieq}) we can
get the equations of perturbations. We can solve them most easily
by decomposing into helicities, using the transverse-traceless
tensor projection operator $K_{\mu \nu \alpha \beta}$, such that
$K_{\mu\nu}{}^{\alpha \beta} h_{\alpha \beta} = \gamma_{\mu\nu}$.
This operator must satisfy $K_{\mu\nu\alpha\beta} =
K_{\nu\mu\alpha\beta} = K_{\mu\nu\beta\alpha}$ and
$K^\mu{}_{\mu\alpha\beta} = \partial^\mu K_{\mu\nu\alpha\beta}=
0$. The unique solution is
\ba K_{\mu\nu\alpha\beta} &=& - \,
\frac{1}{3}\Bigg[\eta_{\mu\nu}\eta_{\alpha\beta}
-\frac{3}{2}\bigg(\eta_{\mu\alpha}\eta_{\nu\beta} +
\eta_{\mu\beta}\eta_{\nu \alpha}\bigg) -
\eta_{\mu\nu}\frac{\partial_\alpha \partial_\beta}{\partial_4{}^2}
-  \eta_{\alpha\beta}\frac{\partial_\mu \partial_\nu}{\partial_4{}^2}
\nonumber \\
&& + \, \frac{3}{2}\bigg( \eta_{\nu \beta} \frac{\partial_\mu
\partial_\alpha}{\partial_4{}^2}+ \eta_{\mu\beta}
\frac{\partial_\nu \partial_\alpha}{\partial_4{}^2}
+ \eta_{\nu \alpha} \frac{\partial_\mu \partial_\beta}{\partial_4{}^2}
+ \eta_{\mu\alpha} \frac{\partial_\nu \partial_\beta}{\partial_4{}^2} \bigg)
- 2\frac{\partial_\mu \partial_\nu \partial_\alpha
\partial_\beta}{\partial_4{}^4}\Bigg] \, . \label{projop} \ea
Since the vacuum (\ref{thick6dvac}) is $4D$-flat, stress energy
conservation implies $\partial_\mu \tau^\mu{}_\nu = 0$, and so
$K_{\mu\nu}{}^{\alpha\beta}\tau_{\alpha\beta} = \tau^\mu{}_\nu -
\frac{1}{3}\bigg(\delta^\mu{}_\nu - \frac{\partial^\mu
\partial_\nu}{\partial_4{}^2}\bigg) \tau^\alpha{}_\alpha$.
Hence the {\tt TT}-tensor field equation is
\be \gamma^\mu{}_\nu{}'' + \frac{\alpha'}{2\alpha}
\gamma^\mu{}_\nu{}' - k^2 \gamma^\mu{}_\nu - \frac{M_5^3}{M_6^4}
k^2  \gamma^\mu{}_\nu \, \delta(\rho-r_0) =
-\frac{2}{M_6^4}\bigg[\tau^\mu{}_\nu -
\frac{1}{3}\bigg(\delta^\mu{}_\nu - \frac{k^\mu k_\nu}{k^2}\bigg)
\tau^\alpha{}_\alpha \bigg]\delta(\rho-r_0) \, . ~~~
\label{tttensorsa} \ee

Subtracting this from the full set of equations isolates the
vectors and scalars. Because the backgrounds (\ref{thick6dvac})
are Lorentz-invariant, the vector sources vanish at the linear
order and so the vectors decouple from the matter distribution, so
they can all be set to zero. The scalars remain. They are all
controlled by a single independent scalar field, the `radion',
defined by $X = \frac{1}{4} (h_4 - \partial_4{}^2 \Psi)$, where
$\Psi$ is fully determined by $X$ \cite{kkland}, and ${}^s
h^\mu{}_\nu = \partial^\mu \partial_\nu \Psi + X\delta^\mu{}_\nu$.
A straightforward albeit lengthy calculation \cite{kkland} then
yields the field equation for it, which can be written as
\ba && X'' + \frac{\alpha'}{2\alpha} X' - k^2 X -
\frac{M_5^3}{M_6^4} \bigg(k^2 + \frac{3b M_6^4}{4M_5^3r_0}  -
\beta(k^2) \bigg) X \, \delta(\rho-r_0) =
~~~~~~~~~~~~~~~~~~~~~~~~~~~~ \nonumber \\
&& ~~~~~~~~~~~~~~~~~~~~~~~~~~~~~~~ - \frac{1}{6M_6^4} \,
\Bigg({\tau}^\alpha{}_{\alpha} + \,  \frac{3M_5^3 r_0 k^2}{b M_6^4}
\, \, \frac{1-b - \frac{b M_6^4 r_0}{M_5^3}}{1-b
+k^2 r_0^2} \,{\tau}^\phi{}_{\phi} \Bigg) \, \delta(\rho-r_0) \, ,
\label{xxaf} \ea
where
\be
\beta(k^2) =  \frac{3M^3_5 r_0}{4 b M_6^4} \,
\frac{1-b}{1-b + k^2 r_0^2} \, \bigg(k^2 +
\frac{b M_6^4}{M_5^3 r_0}\bigg)^2 \, , \label{brscals}
\ee
is the correction to the brane kinetic term of the scalar $X$
originating from the brane bending contributions. This equation is
very similar to the {\tt TT}-tensor equation (\ref{tttensorsa}),
with the differences being in the brane-localized terms.

We can solve both of these equations using the standard Green's
function techniques \cite{kkland}. The Green's functions for the
{\tt TT}-tensor and radion are defined by, respectively,
\ba && {\cal G}_{\tt TT}{}'' + \frac{\alpha'}{2\alpha} {\cal G}_{\tt
TT}{}' - k^2 {\cal G}_{\tt TT}  - \frac{M_5^3}{M_6^4} k^2 {\cal
G}_{\tt TT} \, \delta(\rho-r_0) = \frac{1}{M_6^4} \,
\delta(\rho-r_0) \, , ~~~~~ \label{greens} \\
&& {\cal G}_{\tt X}{}'' + \frac{\alpha'}{2\alpha} {\cal G}_{\tt
X}{}' - k^2 {\cal G}_{\tt X}  - \frac{M_5^3}{M_6^4} \bigg(k^2 +
\frac{3b M_6^4}{4M_5^3r_0}  - \beta(k^2) \bigg) \, {\cal G}_{\tt X}
\, \delta(\rho-r_0) = \frac{1}{M_6^4} \, \delta(\rho-r_0) \, .
\label{Xgreens} \ea
The solutions are, formally,
\ba \gamma^\mu{}_\nu(k, \rho) &=& - 2 \, {\cal G}_{\tt TT}(k, \rho) \,
\bigg[ \tau^\mu{}_\nu - \frac{1}{3}\bigg(\delta^\mu{}_\nu -
\frac{k^\mu k_\nu}{k^2}\bigg) \tau^\alpha{}_\alpha \bigg] \, ,
\label{grttsoln} \\
X(k, \rho) &=& - \frac16 \, {\cal G}_{\tt X}(k, \rho) \,
\Bigg[{\tau}^\alpha{}_{\alpha} + \,  \frac{3M_5^3 r_0 k^2}{ b M_6^4}
\, \, \frac{1-b - \frac{b M_6^4 r_0}{M_5^3}}{1-b
+k^2 r_0^2} \,{\tau}^\phi{}_{\phi} \Bigg] \, . \label{Xsoln} \ea

Since the theory has scalars, in general one must worry about
which frame one extracts the long range fields in. The long
distance dynamics will be described by a scalar-tensor theory,
where the radion is the Brans-Dicke-like scalar. To go to the
effective `Einstein' frame, we can conformally remove the radion
which on the brane yields the Einstein frame metric perturbation
\be h^{\tt E}{\,}^\mu{}_\nu = \gamma^\mu{}_\nu -  4 \, \bigg( X -
\frac{r_0 k^2}{2} \xi \bigg) \, \frac{k^\mu k_\nu}{k^2} + \bigg(
\frac{\xi}{r_0} - \frac{X}{2} \bigg) \, \delta^\mu{}_\nu\, ,
\label{hsolnfef} \ee
where of course $\xi$ is given by $X$ \cite{kkland}. However in
the near-critical limit, it turns out that the radion is
suppressed and so in fact the metric coincides with the {\tt
TT}-tensor to the leading order, as we are about to see.

For static solutions, the Fourier transforms all have $k^\mu = (0,
\vec k)$, and therefore are given by ${\cal G}(k) = 2\pi
\delta(k^0) \, {\cal G}(\vec k)$. Factorizing out the
energy$\delta$-function, setting $k^0=0$ in the formulas of the
previous section, and dropping the integration $\int
\frac{dk^0}{2\pi}$ from the Fourier integral we then solve the
Green's function equations (\ref{greens}) and (\ref{Xgreens}). On
the brane $\rho = r_0$ at distances $\ell \gg r_0$ we must be
cautious with $K_0$ and $K_1$ since they depend on the deficit
angle, as in the shock wave case. The consistent approximation for
the Green's functions on the brane is \cite{kkland}
\ba {\cal G}_{\tt TT} &=&  \frac{1}{M_6^4 } \,
\frac{1}{\frac{K_1(k\frac{r_0}{1-b})}{K_0(k\frac{r_0}{1-b})} k +
\frac{M_5^3}{M_6^4} k^2 }
\, ,   \label{ttgrsolnbr} \\
{\cal G}_{\tt X} &=&  \frac{1}{M_6^4} \,
\frac{1}{\frac{K_1(k\frac{r_0}{1-b})}{K_0(k\frac{r_0}{1-b})} k
+\frac{M_5^3}{M_6^4} [k^2 + \frac{3bM_6^4}{4M_5^3 r_0} -
\beta(k^2)]} \, ,   \label{Xgrsolnbr} \ea
For the {\tt TT}-tensor, the behavior of the theory is governed by
the ratio $k_c = \frac{M_6^4 K_1(k\frac{r_0}{1-b})}{M_5^3
K_0(k\frac{r_0}{1-b})}$, which controls the denominator of ${\cal
G}_{\tt TT}$ at low momenta. For $k < k_c$, or distances $\ell >
1/k_c$, the dominant contributions always come from $\propto
\frac{K_1(k\frac{r_0}{1-b})}{K_0(k\frac{r_0}{1-b})}$. In this
limit the long range fields manifestly reveal the extra
dimensions, since the scaling of the potentials will not be $4D$.
The critical value $k_c$ is {\it exactly} the same as the one
discovered by the shock wave analysis. Thus specifically in the
near-critical limit, the crossover scale is $r_c =
\frac{M^3_5}{M_6^4}$, where gravity first changes into a $5D$
theory. Eventually, the circle opens up and the theory turns into
a $6D$ gravity on a cone. The crucial dynamics which manufactures
this sequence of transitions is the compactification of one bulk
dimension induced by a near-critical brane \cite{kalwall}.

The function ${\cal G}_{\tt X}$ is considerably more involved
because of $\beta(k^2)$ in the denominator. It behaves differently
for generic sub-critical and for near-critical branes. For generic
sub-critical branes, $b \la 1$, at scales $k \ll 1/r_0$, ${\cal
G}_{\tt X}$ function (\ref{Xgrsolnbr}) reduces to
\be
{\cal G}_{\tt X} = \frac{1}{M_6^4} \,
\frac{{r_0\ln[\frac{2(1-b)}{kr_0}]} }{{1-b} - \frac{3 M_5^6r^2_0}{4b M_6^8}
k^2 \bigg( k^2 + \frac{2bM_6^4}{3M_5^3 r_0}\bigg) {\ln[\frac{2(1-b)}{kr_0}]} } \, ,
\label{Xgrappr}
\ee
where we have approximated
$\frac{K_1(k\frac{r_0}{1-b})}{K_0(k\frac{r_0}{1-b})}$ with the
small argument expansion. At very large distances ${\cal G}_{\tt
X} \simeq \frac{{r_0\ln[\frac{2(1-b)}{kr_0}]} }{{(1-b)}M_6^4}$,
implying that the configuration space solution for the scalar
field $X$ depends on the distance as $X \sim {\cal M}/|\vec x|^3$
- i.e. as a field in $6D$. But as we get closer to the source, the
negative momentum-dependent terms in the denominator of ${\cal
G}_{\tt X}$ make the scalar force grow, and the scalar eventually
becomes strongly coupled as $k$ approaches the pole of
(\ref{Xgrappr}). This happens when $k \sim k_*$, where
\be k^2_* \simeq \frac{2(1-b)M_6^4}{M_5^3 r_0
\ln[\frac{2(1-b)}{kr_0}]} \, . \label{strcoupl} \ee
But this is basically the same as the crossover momentum in the
{\tt TT}-tensor sector! So for sub-critical branes inside the
regime of length scales $\ell < r_*$ where
\be
r_*^2 \simeq \frac{M_4^2} {2(1-b)M_6^4}  \ln[\frac{2(1-b)}{kr_0}] \, ,
\label{strongcoupllength}
\ee
the scalar sector is in fact strongly coupled. This makes the
perturbation theory around the vacuum on sub-critical branes
completely unreliable inside the regime where gravity might be
$4D$. The negative signs of the momentum-dependent terms in the
denominator of (\ref{Xgrappr}) might naively suggest presence of
ghosts in this regime, but at this level of the approximation we
cannot conclude that decisively. We conclude that at the level of
linearized perturbation theory the gravitational effects below the
crossover scale are {\it not} calculable for generic subcritical
cases.

In the near-critical limit the scalar sector behaves very
differently. At distances $\ell \gg r_0$, the scalar Green's
function is \cite{kkland}
\be
{\cal G}_{\tt X} = \frac{1}{M_6^4} \,
\frac{1}{k +\frac{M_5^3}{M_6^4}
 \frac{r_0^2 k^2}{1-b + r_0^2 k^2}(k^2
+ \frac{3}{4r_c r_0} )} \, ,\label{Xgrapprext1} \ee
where $r_c = M^5_3/M_6^4$ is the {\tt TT}-tensor crossover scale.
Now, it is convenient to define the scale $r_{\tt vac} =
r_0/\sqrt{1- b}$. When $b \rightarrow 1$ we have $r_{\tt vac} \ll
\frac{r_0}{1-b}$, and so the scale $r_{\tt vac}$ is always smaller
than the size of the conical throat surrounding the brane. This is
why the scale $r_{\tt vac}$ may compete with $r_c$ for the control
over the scalar sector. Both $r_{\tt vac}$ and $r_c$ are smaller
than the size of the throat, and the details of sub-crossover
dynamics depend on their ratio. Skipping some details, we can
check \cite{kkland} that the scalar $X$ is either more weakly
coupled to matter than the {\tt TT}-tensor (when $r_{\tt vac} <
r_c$) or is Yukawa-suppressed, by a mass term $m^2_{\tt X} =
\frac{3}{4r_c r_0}$ (when $r_{\tt vac} > r_c$). Either way, the
long range forces generated by scalars in the linearized
approximation on near-critical branes turn out to  be small
compared to those arising from the {\tt TT}-tensors.

Therefore the effective $4D$ theory below the crossover scale
$r_c$, in the regime of distances $r_0 \ll \ell \ll r_c$ is
completely controlled by the {\tt TT}-tensors. In this regime
their Green's function reduces to ${\cal G}_{\tt TT} \rightarrow
\frac{1}{M_5^3 k^2 }$. For non-relativistic matter sources
$\tau^a{}_b = - \mu \, {\rm diag}(1,0,0,0,1)$ the linearized
solution is
\be \gamma^\mu{}_\nu(k, \rho) = \frac{2 \mu}{3 M_5^3 k^2} \,
\cases{ 2 \, , ~~~~~ & $\mu = \nu = 0 \,$,\cr
 - \bigg(\delta^j{}_k -
\frac{k^j k_k}{k^2}\bigg) \, , ~~~~~ & $\mu = j, \, \nu = k \,$.}
\label{ttsolstat}
\ee
Fourier-transforming to coordinate space and recalling that $\mu$
is the mass per unit length of string, so that $\mu = \frac{{\cal
M}}{2\pi r_0}$, and using as before the Gauss law $M_5^3 =
\frac{M^2_4}{2\pi r_0}$, and introducing the effective $4D$
Newton's constant $G_{N \, eff} = \frac{1}{8\pi M_4^2}$ we finally
obtain the {\tt TT}-tensor in the $4D$ regime:
\ba \gamma^0{}_0(\vec x) &=&  \frac83 \, G_{N \, eff}
\, \frac{\cal M}{ |\vec x|} \, , \nonumber \\
\gamma^j{}_k(\vec x) &=&  - \frac23 \, G_{N \, eff} \, \frac{\cal M}{ |\vec x|} \,
\Bigg(\delta^j{}_k + \frac{x^j x_k}{\vec x^2}  \Bigg)
\, .
\label{gammaall}
\ea
The Newtonian potential $V_N = - h^0{}_0/2 = -\frac43 \, G_{N \,
eff} \, \frac{\cal M}{ |\vec x|}$, is a factor of $4/3$ larger
than the usual formula of General Relativity, which a
manifestation of the Iwasaki-van Dam-Veltman-Zakharov
discontinuity \cite{vdvz}. This signals the presence of the
helicity-0 modes in the graviton multiplet. The factor $4/3$
enhancement is precisely what one expects based on other examples
\cite{DGP}, and in this case shows that the extra helicity-0 modes
in the spin-2 multiplet mediate attractive force, just as the
helicity-2 modes.  In fact the solution (\ref{gammaall}) mimics a
Brans-Dicke theory with $\gamma = \frac12$, or therefore, with the
Brans-Dicke $\omega$ parameter equal to zero, where the parameter
$\omega$ is defined in the usual way in the Brans-Dicke action as
$S_{BD} = \int d^4x \sqrt{g} (\Phi R - \omega (\nabla
\Phi)^2/\Phi)$. This theory is in conflict\footnote{Ideas for decoupling of the Brans-Dicke
scalar by using radiatively induced potential were explored in \cite{cliff}, however the field $\Phi$ here
is, as we noted, really the helicity-0 graviton.} with observational data
as it stands, and hence is not a realistic description of our
Universe.

However, it is conceivable (although not certain) that different
strong coupling effects at higher orders in perturbation theory
might improve the behavior of long range fields of masses far from
the source but well below the crossover scale
\cite{veinsh,strongcouplings}-\cite{nira}. Another possibility
might be that the additional tweaks of the bulk theory, e.g.
curving the bulk locally, could change how the helicity-0 mode
couples to brane matter. Either way, the scalar couplings remain
consistently {\it weak} for static sources on near-critical branes
with $b \rightarrow 1$. The static solutions are governed by
Euclidean momenta, and so they are always finite in linearized
theory around near-critical vacua, in contrast to the fields of
masses which perturb generic sub-critical vacua. Thus linearized
perturbation theory remains under control on near-critical vacua.
Similar arguments show that the near-critical propagating
solutions  also remain under control \cite{kkland}.

\section{Summary}

BIG models may in fact approximate $4D$ worlds between the UV
cutoff that resolves the core of the gravity-localizing defect and
the crossover scale, if their Scherk-Schwarz sector is carefully
tuned to yield $4D$ Minkowski vacua. We have seen this explicitly
in the example of regulated codimension-2 defects, which remain
perturbatively calculable and have a $4D$ regime for near-critical
tensions. We feel that similar conclusions may extend to any
setups with codimension $\ge 2$. The gravitational sector contains
additional scalars, but at least at the linearized order there are
no ghosts, of the kind that plague the codimension-1 DGP models
\cite{Luty,koyama}-\cite{minjoon}, or were argued to arise in soft
gravity resolutions of higher-codimension cases \cite{durub}.

The cosmological constant problem changes completely in BIG
codimension-2 setups. Putting brane localized terms and changing
gravity in the IR by itself does not solve it. Instead, one still
has vacuum energy problem, since the tension of the brane needs to
be tuned to get a desired low energy theory. But now, vacuum
energy does not need to curve the $4D$ geometry at all. Instead,
we must tune the vacuum energy to get a weak $4D$ gravity over a
large spatial region and maintain perturbativity in the scalar
graviton sector. This might offer a different strategy to help
with the vacuum energy problem: perhaps the fact that graviscalars
go into the strong coupling for detuned tensions can be used as a
new tool. One might try to look, for example, for theories where
the tuned value of the tension, which ensures perturbativity,
marks an IR fixed point of the combined matter and graviscalar
theory. In this case, at least some of the premises of the
Weinberg's no-go theorem for cosmological constant adjustment
\cite{Wein} would be violated, and it would be interesting to see
what the consequences are.

The numerology of the near-critical codimension-2 BIG is rather
intriguing. Using the formula for the $4D \rightarrow 5D$
crossover scale $r_c = M_5^3/M_6^4$, we see that to get $r_c$ to
be of the order of the current horizon size, $H_0^{-1} \sim
10^{28} {\rm cm}$, and ensure that matter on the brane looks
pointlike at energies below a ${\rm TeV}$, we need $M_6 \ga {\rm
TeV}$, and therefore $M_5 \la 10^{19} {\rm GeV}$. Further to
reproduce the $4D$ Newton's constant below the crossover scale,
one needs to pick $r_0 \ga 10^{-19} {\rm GeV}^{-1}$. So we must
tune the theory to make sure that the brane is very thin, with
many hidden sector fields, to get such hierarchies. We don't have
much to add to the discussion of how to do it, but merely note
that this is in line with the current philosophy of the BIG 
models \cite{DGP}. While some work along the lines
of embedding the theory into string theory, that would provide the
framework for its full UV completion, has been pursued
\cite{kiritsis,lowe,ignatios}, the existing constructions are
really semiclassical models where gravity is treated classically.
At least, we can explore the low energy consequences of such
models.

Many questions remain open. Can the helicity-0 modes decouple? Are
there dangerous instabilities beyond the linear order? Can BIG
theory be UV-completed? Most interestingly, could the new guise of
the cosmological constant problem open some new directions to
circumvent the Weinberg no-go theorem \cite{Wein}? While we can't
address these issues here, we hope to return to them soon.

\vskip.4cm

{\bf \noindent Acknowledgements}

\vskip.2cm

We would like to thank Derrick Kiley for collaboration and 
Alberto Iglesias, Robert Myers, Minjoon Park, Oriol Pujolas, Lorenzo Sorbo
and John Terning for interesting discussions. N.K. is grateful to
Galileo Galilei Institute, Florence, Italy, for kind hospitality
in the course of this work, and to the organizers of the ``Sowers
Workshop", Virginia Tech, May 14-18, 2007, ``Cosmology and
Strings" workshop at ICTP, Trieste, Italy, July 9-13, 2007, ``Dark
Energy In the Universe", Hakone, Japan Sep 1-4, 2007 \cite{hakone}
and ``Zagreb Workshop 2007", Zagreb, Croatia, Nov 9-11, 2007, for
the opportunity to present the results and the kind hospitality.
This work has been supported in part by the DOE Grant
DE-FG03-91ER40674, in part by the NSF Grant PHY-0332258 and in
part by a Research Innovation Award from the Research Corporation.


\end{document}